%% file: main.tex
\documentclass{article}
\usepackage{spconf,amsmath,graphicx}
\usepackage{booktabs,array,cite,url}
\usepackage{microtype,needspace,afterpage}
\usepackage{stfloats}
\input{config/metadata.tex}

\makeatletter
\def\ps@preprintfirst{%
  \def\@oddhead{}\def\@evenhead{}%
  \def\@oddfoot{\hfil\parbox[b]{0.90\textwidth}{\centering\fontsize{6}{7}\selectfont
  This work has been submitted to the IEEE for possible publication. Copyright may be transferred without notice, after which this version may no longer be accessible.}\hfil}%
  \let\@evenfoot\@oddfoot}
\makeatother

\begin{document}
\maketitle
\thispagestyle{preprintfirst}

% Avoid stretched section spacing on the first page.
% Restore aligned column bottoms from page 2 onward.
\raggedbottom
\afterpage{\flushbottom}

\input{sections/00_abstract.tex}
\input{sections/01_introduction.tex}
\input{sections/02_data_and_interfaces.tex}
\input{sections/03_experimental_protocol.tex}
\input{sections/04_results.tex}
\input{sections/05_discussion.tex}

% ISBI permits five pages when the fifth page is limited to ethical compliance,
% acknowledgments/disclosures, and references.
\clearpage
\raggedbottom
\input{sections/06_ethics.tex}
\input{sections/07_acknowledgments.tex}

\bibliographystyle{IEEEbib}
\bibliography{references}
\end{document}

%% file: config/metadata.tex
\title{Supervision-Aware Evaluation of MedGemma for
Reference-Assisted Radiotherapy Contour Ranking}

\name{
Jaden Moon\textsuperscript{1,2} \qquad
Caiwen Jiang\textsuperscript{2} \qquad
Songlin Zhao\textsuperscript{2} \qquad
Wei Liu\textsuperscript{2}
}

\address{
\textsuperscript{1}Dartmouth College, Hanover, NH, USA\\
\textsuperscript{2}Mayo Clinic, Phoenix, AZ, USA
}

%% file: sections/00_abstract.tex
\begin{abstract}
Radiotherapy contour ranking can support quality assurance, but
vision--language model (VLM) gains may reflect supervision, score
construction, available inputs, or representation access. We evaluated frozen
MedGemma~1.5 in a single-institution study, a 12-patient held-out HaN-Seg
test, fivefold cross-fitting on 30 non-test HaN-Seg patients, and transfer
without refitting to 120 SegRap2023 patients. On held-out HaN-Seg, a
supervised readout of the final prompt state exceeded a supervised readout of
the model's yes/no output probabilities by \(+0.3047\) pooled average
precision (AP; 95\% patient-bootstrap interval
\([+0.2822,+0.3269]\)); the advantage persisted against a
parameter-count-matched nonlinear control. A readout of projected visual
tokens exceeded the final-state readout in cross-fitting and transfer
(\(+0.0706\) AP averaged equally across patients on SegRap; 95\% interval
\([+0.0607,+0.0804]\)), while a structured panel-feature readout remained competitive with visual
tokens. The anatomy-conditioned readout did not demonstrate an advantage over the
native-feature comparator. These results motivate visual-token and
structured-feature controls in medical-imaging VLM evaluations; outcomes were
computational or synthetic, not clinical acceptability.
\end{abstract}


%% file: sections/01_introduction.tex
\section{Introduction}

\input{figures/figure1.tex}

\begingroup
\setlength{\parskip}{0pt}

Organ-at-risk (OAR) contour quality assurance (QA) uses segmentation-based
checks, learned quality predictors, and vision--language models (VLMs)
\cite{men2020oarqa,isaksson2022contourqa,luan2024vlmcontourqa}. When a VLM
score improves ranking, the gain may reflect supervision, score mapping,
representation access, or displayed information; controlled comparisons are
therefore needed before attributing it to the pretrained representation. We ask
whether internal features add value beyond supervised remapping of output
probabilities.

Probing depends on both features and fitted scorers
\cite{hewitt2019probes,belinkov2022probes}. MedProb finds signal beyond direct
prompting \cite{nourbakhsh2026medprob}, a dermatology study reports stronger
visual-encoder signals than the full VLM \cite{wang2026underutilized}, and
SegAE assesses segmentation-label quality \cite{chen2026segae}. These studies leave internal-versus-output-space scoring unresolved in
reference-assisted radiotherapy contour ranking.

Figure~\ref{fig:method} maps the institutional (A) and public (B) scoring
interfaces---feature sources paired with scoring rules---for frozen
MedGemma~1.5 \cite{sellergren2026medgemma15}. The Mayo study compares an
unfitted token score, supervised final-state readouts, and a native-feature model;
HaN-Seg adds patient-disjoint testing and non-test controls, followed by
source-fitted readouts applied unchanged to SegRap2023
\cite{podobnik2023hanseg,luo2023segrap2023}.

We contribute: (1) controlled comparisons of task fitting, score mapping,
representation access, and displayed information while making residual input
differences explicit; (2) patient-disjoint testing and transfer without
external refitting; and (3) visual-token and panel-feature controls showing
that the observed gain is not unique to the final prompt state. Because
reference geometry is available, we evaluate MedGemma as a scoring interface,
not as a replacement for direct geometry or clinical review.

\endgroup

%% file: figures/figure1.tex
\begin{figure*}[t]
\centering
\includegraphics[width=\textwidth]{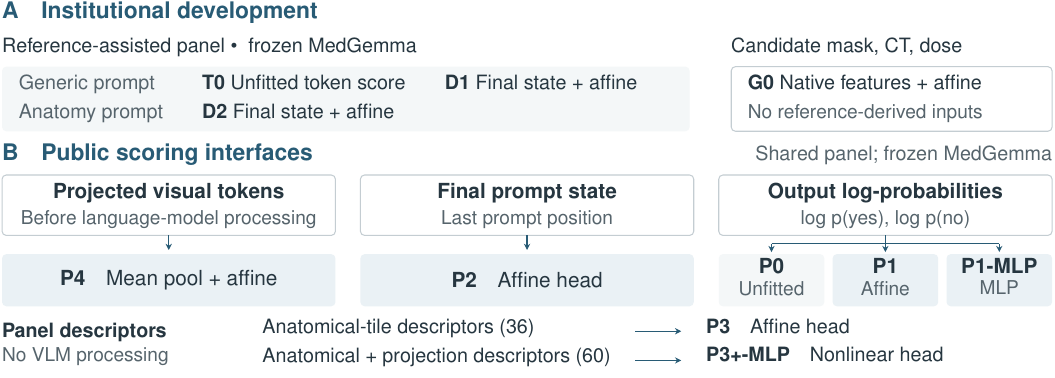}
\caption{\textbf{Scoring interfaces.}
(A) Institutional paths compare reference-assisted T0/D1/D2 with native-input
G0; (B) public paths compare output-space, final-state, visual-token, and
structured-panel readouts. T0/P0 are unfitted; other heads are task-trained.
P0/P1/P2/P3 use the 12-patient test; later controls use 30 non-test patients;
separate source fits transfer to 120 SegRap patients.}
\label{fig:method}
\end{figure*}

%% file: sections/02_data_and_interfaces.tex
\section{Methods}
\label{sec:methods}

\begingroup
\setlength{\parskip}{0pt}

\newcommand{\methodblock}[1]{%
  \par\addvspace{0.25\baselineskip}%
  \noindent\textbf{#1}\nobreak\hspace{0.4em}\ignorespaces
}

\subsection{Institutional study and computational target}
\label{sec:task-data}

\methodblock{Study population.}
Mayo Clinic Arizona computed tomography (CT) images and consolidated
multi-label reference masks were resized to \(256\times256\times128\).
Rows represented candidate OARs with reference and dose context; recorded
identifiers defined case groups. Training contained 33,260 rows, 832 positives,
and 676 groups; development contained 3,968 rows, 66 positives, and 76 groups,
with 23 positives in one group. Identifiers and panels did not overlap across
partitions, but biological-patient separation was unverified.

\methodblock{Computational endpoint.}
The target marks computational discrepancy, not clinical review need.
Training-based selection retained 40 of 42 measurements: dose-endpoint
discrepancy (15), local dose gradient (7), high-dose-shell involvement (7),
shell proximity (8), and candidate--reference disagreement (3).
For candidate \(i\) and measurement \(j\), \(g_j\) orients the recorded
value \(v_{ij}\) so larger
\(u_{ij}=g_j(v_{ij})\) means greater discrepancy.
Let \(q_{j,a}\) be the training \(a\)th percentile of \(u_{ij}\), \(m_j=q_{j,50}\)
the median, and \(r_j=q_{j,75}-q_{j,25}>0\) the interquartile range:
\begin{equation}
\begin{aligned}
c_{ij}&=\left[\frac{\operatorname{clip}(u_{ij},q_{j,1},q_{j,99})-m_j}{r_j}\right]_+,
\\
s_i&=\max_{f\in F_i}\frac{1}{|A_{if}|}\sum_{j\in A_{if}}c_{ij},
\\
y_i&=\mathbf{1}\!\left\{\operatorname{finite}(s_i)\land s_i\geq\tau\right\}.
\end{aligned}
\label{eq:computational-endpoint}
\end{equation}
Here, \(c_{ij}\) is a standardized contribution, \(s_i\) the maximum family
mean, and \(y_i\) the binary label. Clipping bounds values by the supplied
quantiles; \([z]_+=\max(0,z)\), and \(\mathbf{1}\) is an indicator.
\(A_{if}\) indexes row \(i\)'s finite measurements in family \(f\),
\(|A_{if}|\) counts them, and \(F_i\) contains evaluable families.
The threshold \(\tau\approx28.4223\) is the training 97.5th percentile of \(s_i\).
Missing measurements were omitted; rows with no evaluable family
received \(y_i=0\).
All 1,167 empty-mask development rows received the negative computational label. Complete raw measurement
definitions and units were unavailable.

\subsection{Institutional interfaces}
\label{sec:score-interfaces}

\methodblock{Inputs and prompts.}
Figure~\ref{fig:method}A maps institutional paths using candidate/reference
contours, disagreement, and rendered dose after mask-guided cropping and slice
selection. T0, D1, and D2 used
frozen \path{google/medgemma-1.5-4b-it}. T0 and D1 shared the generic
clinical-review prompt \(p_G\); D2 added recorded anatomy and laterality.
Prompts asked about clinical review, whereas readouts predicted computational
discrepancy; these targets were not assumed equivalent.

\methodblock{Scoring interfaces.}
Let \(p_\theta\) be the next-token probability under frozen model parameters
\(\theta\). For panel \(x\) and generic prompt \(p_G\), T0 used
\[
s_{\mathrm{T0}}(x)=\log p_\theta(\texttt{yes}\mid x,p_G)-
\log p_\theta(\texttt{no}\mid x,p_G),
\]
without task-label fitting. D1 and D2 fitted affine readouts (linear scores
with bias) to the 2,560-dimensional final state at the last prompt position
under the generic and anatomy-conditioned prompts. G0 used the same head class on 41 candidate-mask, CT, and dose features,
excluding reference-derived inputs, anatomy identity, metadata, and
outcome-derived fields.

\methodblock{Comparison rationale.}
D1--T0 is the closest institutional control: it matches panel and generic
prompt but jointly changes supervision, feature access, and scoring. D2--T0
also changes anatomy conditioning. D2--G0 compares unequal-input pipelines,
not representations alone.

\subsection{Public patient-disjoint study}
\label{sec:public-study}

\methodblock{HaN-Seg candidates and panels.}
We split 42 HaN-Seg Set~1 patients \cite{podobnik2023hanseg} deterministically
into 22 training, 8 validation, and 12 test patients. Eight OARs were fixed
using training-only presence, volume, geometry, perturbation feasibility, and
size quantiles, then retained for cross-fitting. Six perturbation
families (three translation axes, erosion, dilation, and principal-axis scaling)
formed three-candidate patient--OAR--family \emph{blocks}. Mild, moderate,
and severe candidates used fractions \(\alpha=0.05,0.10,0.20\), respectively.
On 1-mm isotropic grids, translation distances were \(\alpha\) times the reference bounding-box diagonal
(clipped to 1--8\,mm); erosion/dilation radii were \(\alpha\) times the
radius of a sphere with the reference volume (clipped to 0.5--3\,mm).
Scaling used \(1\pm\alpha\) along the longest principal axis about the
reference centroid. Directions were fixed within each block. Severe candidates were labeled 1; the others were labeled 0. Blocks containing an empty, nonfinite,
nonbinary, or grid-mismatched mask were excluded, leaving 6,027 rows:
3,159 training, 1,149 validation, and 1,719 test.

\noindent\textbf{Panel rendering.}\hspace{0.4em}
CT and expert reference masks were rendered in lossless \(896\times896\)
red--green--blue (RGB) panels using a fixed CT window, a 20-mm margin around
the candidate/reference union,
maximum-disagreement axial, coronal, and sagittal slices, and three orthogonal
disagreement projections. Panels contained no patient, anatomy, perturbation,
severity, or outcome text.

\needspace{3\baselineskip}
\methodblock{SegRap cohort.}
External evaluation used noncontrast CT and updated Task001 masks from all 120 labeled patients in the
SegRap2023 training set \cite{luo2023segrap2023}.
Source fitting and external evaluation used five prespecified OARs shared
with the selected HaN-Seg task: right submandibular gland, spinal cord, thyroid, left
parotid, and oral cavity. Perturbations, labels, rendering, prompt, and
feature definitions were unchanged.

% Keep the interface heading with its opening explanation.
\needspace{3\baselineskip}
\methodblock{Public interfaces.}
Figure~\ref{fig:method}B summarizes the public paths; P0 used the unfitted
yes/no log-probability difference under the prompt:
``A candidate organ-at-risk contour and its reference are shown. Does the
candidate have a severe geometric discrepancy? Answer only yes or no.''
P1 fitted an affine head to the two output log-probabilities; P2 fitted the
same head class to the 2,560-dimensional final state at the last prompt
position. P3 used an affine head on 36 anatomical-tile descriptors: area,
overlap, centroid, connectivity, boundary distance, and CT summaries.
It excluded projection tiles.

\noindent\textbf{Follow-up controls.}\hspace{0.4em}
Fitting used only non-test HaN-Seg patients. P1-MLP used
a multilayer perceptron (MLP; \(2\!\to\!640\!\to\!1\), 2,561 parameters) to test nonlinear output remapping
at P2's parameter count, without assuming equal functional capacity.
P3+ denotes the addition of 24 projection descriptors to P3's 36 descriptors
(60 total): disagreement area, bounding-box extent/occupancy, components,
boundary density/ratio, and eccentricity. P3+-MLP replaced P3's affine head with an MLP
(\(60\!\to\!41\!\to\!1\), 2,543 parameters) on these 60 features. P4 used an affine head on the mean of
256 visual tokens after the image projector (2,560 dimensions). P4--P2
matches panels, checkpoint, dimension, head class, folds, weights, and
optimization, but changes extraction, token selection, and pooling.

\endgroup

% Keep the frozen result tables; adjust only the surrounding float separation.
\setlength{\dbltextfloatsep}{10pt plus 2pt minus 2pt}
% The table inputs are queued in Section 2.4 to retain the results-page layout.

%% file: sections/03_experimental_protocol.tex
\subsection{Training, evaluation, and controls}
\label{sec:training-statistics}

\begingroup
\setlength{\parskip}{0pt}

\newcommand{\protocolblock}[1]{%
  \par\addvspace{0.25\baselineskip}%
  \noindent\textbf{#1}\nobreak\hspace{0.4em}\ignorespaces
}

Pooled AP combines all rows; patient-macro AP gives each patient's AP equal
weight. We averaged seed-specific metrics, not predictions. Paired 95\%
bootstrap confidence intervals (CIs) used 2,000 institutional case-group
draws and 10,000 patient draws in the public studies. All rows were retained
for each sampled occurrence; models and predictions remained fixed.

\protocolblock{Institutional fitting.}
Affine heads minimized binary cross-entropy on training-standardized features
(standard-deviation floor \(10^{-6}\)) using
AdamW \cite{loshchilov2019decoupled}: learning rate \(10^{-3}\), weight decay
\(10^{-4}\), batch size 1,024, 40 epochs, and seeds 17/43/71.
Rows sharing a panel received total weight one. All final-epoch seeds were retained without selection. AP was primary at 1.663\% prevalence
\cite{saito2015precisionrecall}; resampling used the 76 development case
groups \cite{field2007clusterbootstrap}.

\protocolblock{Held-out HaN-Seg evaluation.}
P1--P3 used the affine protocol with training-only standardization and
patient-balanced weights (equal total weight per patient). Validation labels
did not guide model, feature, seed, or checkpoint choice. Heads were
refitted on training plus validation and tested once on 12 patients. Pooled
AP was the primary metric; P2--P1 and P2--P3 were the primary and key
secondary contrasts. Supporting metrics were area under the receiver
operating characteristic curve (AUROC) and patient-macro AP/AUROC.

\protocolblock{Cross-fitting and zero-refit transfer.}
Each fold used 24 of 30 non-test patients for standardization and fitting,
evaluating P1-MLP, P2, P4, and P3+-MLP on the other six. Transfer used
separate five-OAR readouts fitted on all 30 source patients (2,700 rows),
not the eight-OAR cross-fit models. Both analyses used patient-balanced weights,
80 epochs, and seeds 17/43/71, retaining every final-epoch fit. Transfer
applied source preprocessing, prompts, standardizers, and weights unchanged,
without external fitting or calibration. Patient-macro AP was primary.

\input{tables/original_comparisons.tex}
\input{tables/public_contrasts.tex}

\protocolblock{Panel-identifiable sensitivity.}
The primary SegRap population comprised all 3,600 valid blocks (10,800 rows).
Clipping and rasterization sometimes made distinct candidates render identically with conflicting labels. A fixed rule removed complete blocks containing any panel assigned both
labels anywhere in the cohort. It retained
2,057 blocks (6,171 rows) and all 120 patients, but fewer OAR--perturbation
combinations. This sensitivity does not replace the primary analysis;
higher AP on the restricted population does not establish model improvement.

\protocolblock{Baseline negative controls.}
Cross-fitting used P2/P4 label permutations; transfer used a separate P2
control. Training positives were reassigned within patient--OAR--family blocks,
while evaluation labels remained fixed.

\protocolblock{Geometric-rank sensitivity.}
Without refitting, we compared scores with within-block midranks of
\(1-\mathrm{Dice}\) and HD95, averaging tied ranks. HD95 was the larger
directed 95th-percentile surface distance in millimetres; higher ranks indicated
greater discrepancy. Concordance gave half credit to score ties and was averaged
equally across eligible blocks within patient, then patients and seeds. In 1,000
tie-preserving target permutations, each model mean had to lie in [0.49, 0.51]
with its central 95\% range containing 0.5; each contrast required an absolute
mean of at most 0.01 and a central range containing zero. AP remained primary.

\protocolblock{Held-out-class evaluation.}
Each class---translation, erosion, dilation, or scaling---was excluded from
fitting and evaluated in non-test HaN-Seg and SegRap. P2 controls used 256 severity-balanced label permutations per class. For patient-macro AUROC and
mild \(<\) moderate \(<\) severe concordance, the control mean and 95\% CI
had to lie within [0.49, 0.51] for every class and the equal-weight mean in both
populations.
\endgroup

%% file: tables/original_comparisons.tex
% Original evaluations remain separate tables, sharing one full-width float.
\begin{table*}[!t]
\centering
\begin{minipage}[t]{\dimexpr(\textwidth-\columnsep)/2\relax}
\input{tables/main_results.tex}
\end{minipage}\hfill
\begin{minipage}[t]{\dimexpr(\textwidth-\columnsep)/2\relax}
\input{tables/public_replication.tex}
\end{minipage}
\end{table*}

%% file: tables/main_results.tex
\centering
\caption{\textbf{Institutional development and AP contrasts.}}
\label{tab:results}
\small
\setlength{\tabcolsep}{1pt}
\renewcommand{\arraystretch}{1.0}
% Keep contrast estimates and interval bounds aligned, including negative bounds.
\newcommand{\institutionalci}[3]{%
  \makebox[3.9em][r]{\(#1\)}\;\makebox[7.6em][l]{\([\ifdim#2pt<0pt\else\phantom{-}\fi#2,\,#3]\)}%
}
\begin{tabular*}{\linewidth}{@{\extracolsep{\fill}}lrrr@{}}
\toprule
\textbf{Interface} &
\shortstack{\strut\textbf{Development AP}\\\strut\textbf{[95\% CI]}} &
\textbf{AUROC} &
\shortstack{\strut\textbf{Unique}\\\strut\textbf{panel AP}} \\
\midrule
T0: unfitted token & 0.0195 [0.0066, 0.0492] & 0.5419 & 0.0239 \\
D1: generic prompt & 0.1032 [0.0508, 0.2147] & 0.8077 & 0.1085 \\
D2: anatomy prompt & 0.1394 [0.0710, 0.2661] & 0.8623 & 0.1440 \\
G0: native features & 0.1567 [0.0822, 0.3258] & 0.8359 & 0.1568 \\
\midrule
\multicolumn{4}{@{}l}{\textit{Paired AP differences [95\% CI]}} \\
D1--T0 & \multicolumn{3}{r@{}}{\institutionalci{+0.0838}{0.0288}{0.2034}} \\
D2--G0 & \multicolumn{3}{r@{}}{\institutionalci{-0.0173}{-0.1750}{0.0999}} \\
\bottomrule
\end{tabular*}
\par\smallskip
{\small\raggedright
Development: 3,968 rows (66 positives; 76 groups), including 3,096
unique-panel rows. Learned metrics average three seeds; T0 is fixed. CIs use
2,000 paired group bootstraps with fixed predictions.\par}

%% file: tables/public_replication.tex
\centering
\caption{\textbf{Held-out HaN-Seg test and AP contrasts.}}
\label{tab:public-results}
\small
\setlength{\tabcolsep}{2pt}
\renewcommand{\arraystretch}{1.0}
\begin{tabular*}{\linewidth}{@{\extracolsep{\fill}}lrrr@{}}
\toprule
\textbf{Interface} & \shortstack{\strut\textbf{Pooled AP}\\\strut\textbf{(primary)}} & \textbf{AUROC} &
\shortstack{\strut\textbf{Patient-macro}\\\strut\textbf{AP}} \\
\midrule
P0: unfitted token & 0.3169 & 0.4822 & 0.3268 \\
P1: output-space & 0.3200 & 0.4915 & 0.3328 \\
P2: final prompt state & 0.6247 & 0.7767 & 0.6332 \\
P3: anatomical tiles & 0.5701 & 0.7539 & 0.5767 \\
\midrule
\multicolumn{4}{@{}l}{\textit{Paired pooled-AP differences [95\% CI]}} \\
\multicolumn{2}{@{}l}{P2--P1 (primary)} & \multicolumn{2}{r@{}}{$+0.3047\;[0.2822,\,0.3269]$} \\
\multicolumn{2}{@{}l}{P2--P3 (key secondary)} & \multicolumn{2}{r@{}}{$+0.0545\;[0.0292,\,0.0799]$} \\
\bottomrule
\end{tabular*}
\par\smallskip
{\small\raggedright
Test: 12 patients, 1,719 candidates (573 positives). P1--P3 average three
seeds; P0 is fixed. CIs use 10,000 paired patient bootstraps; AUROC and
patient-macro AP are supporting.\par}

%% file: tables/public_contrasts.tex
\begin{table*}[!b]
\centering
\caption{\textbf{Patient-macro AP and paired contrasts in non-test HaN-Seg cross-fitting and zero-refit SegRap transfer.}}
\label{tab:public-contrasts}
\small
\setlength{\tabcolsep}{3pt}
\renewcommand{\arraystretch}{1.0}
% Align signed estimates, interval brackets, and lower-bound signs without rounding.
\newcommand{\pairedap}[3]{%
  \makebox[3.9em][r]{\(#1\)}\;\makebox[7.6em][l]{\([\ifdim#2pt<0pt\else\phantom{-}\fi#2,\,#3]\)}%
}
\begin{tabular*}{\textwidth}{@{\extracolsep{\fill}}lccc@{}}
\toprule
\textbf{Interface or contrast} & \textbf{HaN-Seg cross-fit} &
\multicolumn{2}{c@{}}{\textbf{SegRap zero-refit transfer}} \\
\cmidrule(l){3-4}
 & & \textbf{Full population (primary)} & \textbf{Panel-identifiable sensitivity} \\
\midrule
\multicolumn{4}{@{}l}{\textit{Absolute patient-macro AP}} \\
P1-MLP: nonlinear output-space & 0.3630 & 0.3708 & 0.4074 \\
P2: final prompt state & 0.6339 & 0.5877 & 0.7266 \\
P4: pooled projected visual tokens & 0.7229 & 0.6583 & 0.8171 \\
P3+-MLP: all-panel structured features & 0.7896 & 0.6702 & 0.8290 \\
\addlinespace[3pt]
\multicolumn{4}{@{}l}{\textit{Paired AP differences [95\% CI]; positive values favor the first interface}} \\
P2--P1-MLP & \pairedap{+0.2709}{0.2549}{0.2868} & \pairedap{+0.2169}{0.2036}{0.2301} & \pairedap{+0.3193}{0.2991}{0.3396} \\
P4--P2 & \pairedap{+0.0890}{0.0758}{0.1024} & \pairedap{+0.0706}{0.0607}{0.0804} & \pairedap{+0.0905}{0.0762}{0.1046} \\
P3+-MLP--P4 & \pairedap{+0.0667}{0.0552}{0.0781} & \pairedap{+0.0119}{0.0004}{0.0234} & \pairedap{+0.0119}{-0.0049}{0.0286} \\
\bottomrule
\end{tabular*}
\par\smallskip
{\small\raggedright
HaN-Seg: 30 non-test patients, 4,308 rows, eight OARs. Separate five-OAR source
fits transfer to 120 SegRap patients (10,800 primary; 6,171 sensitivity rows).
Metrics average three seeds; CIs use 10,000 paired patient bootstraps; the
sensitivity restricts blocks, not models.\par}
\end{table*}

%% file: sections/04_results.tex
% Keep the Results heading and opening paragraph together.
\needspace{10\baselineskip}
\section{Results}
\label{sec:results}

\begingroup
\setlength{\parskip}{0pt}

\newcommand{\resultblock}[1]{%
  \par\addvspace{0.25\baselineskip}%
  \noindent\textbf{#1}\nobreak\hspace{0.4em}\ignorespaces
}

\resultblock{Institutional comparison.}
D1 exceeded T0 under the same panel and generic prompt by \(+0.0838\) AP;
D2--T0 was \(+0.1200\) (95\% CI
\([+0.0530,\allowbreak+0.2487]\)). D2--G0 was \(-0.0173\), so D2 did
not demonstrate an advantage over the native-feature model
(Table~\ref{tab:results}).

\resultblock{Held-out public test.}
P2 exceeded P1 by \(+0.3047\) AP and P3 by \(+0.0545\) AP on the held-out
test (Table~\ref{tab:public-results}).

\resultblock{Cross-fitting and transfer.}
P2 exceeded P1-MLP and P4 exceeded P2 in cross-fitting and both SegRap
populations; on primary SegRap, the differences were \(+0.2169\) and
\(+0.0706\) AP (Table~\ref{tab:public-contrasts}). The P3+-MLP--P4
interval included zero only in the panel-identifiable sensitivity. Baseline P2
label-permutation CIs included 0.5 in both SegRap populations; independent code
reproduced the public results.

\resultblock{Geometric-rank sensitivity.}
On primary SegRap, tie-aware geometric-rank contrasts were \(+0.2718\)
[0.2558, 0.2883] for P2--P1-MLP and \(+0.1004\) [0.0911, 0.1099]
for P4--P2. Permutation controls met chance bounds; geometric ranks matched nominal
severity, so this sensitivity addressed ties and duplicate masks rather than a
distinct clinical outcome.

\endgroup

%% file: sections/05_discussion.tex
\section{Discussion and Conclusion}
\label{sec:discussion}

\begingroup
\setlength{\parskip}{0pt}

On the institutional target, supervised MedGemma readouts outperformed token
scoring but did not demonstrate an advantage over the native-feature model.
Public analyses favored final-state over supervised output-space readouts;
follow-up analyses favored projected visual tokens, while the structured
readout remained competitive with visual tokens. The findings support internal
feature access but not a unique final-state advantage or a demonstrated benefit
from language-model contextualization. Extending prior probing work
\cite{nourbakhsh2026medprob,wang2026underutilized}, this study provides a
control hierarchy for medical-imaging VLM evaluation: supervised output-space,
visual-token, and panel-feature comparators should precede
representation-specific claims.

Because reference geometry is available, the contribution is methodological,
not a claim that MedGemma should replace direct measurement or clinical review.
Limitations include prompt--target mismatch, unverified institutional patient
linkage, unequal comparator inputs, synthetic outcomes, reused perturbations,
no calibration, failed held-out-class controls, and unknown pretraining
exposure. Future work should test natural model-generated contours and
clinician-adjudicated outcomes under patient-disjoint evaluation.

\endgroup

%% file: sections/06_ethics.tex
\section{Compliance with Ethical Standards}

Retrospective use of the institutional source cohort was approved by the Mayo
Clinic Institutional Review Board under IRB No.~24-011106, with a waiver of
study-specific written informed consent. The secondary computational analysis
was separately approved under Mayo Clinic IRB No.~24-011064. The study
followed the Declaration of Helsinki. The public HaN-Seg and SegRap2023
analyses involved no additional participant interaction or data collection.

%% file: sections/07_acknowledgments.tex
\section{Acknowledgments}
\label{sec:acknowledgments}

\paragraph*{Funding.}
J.M. was supported by the Mayo Clinic Summer Undergraduate Research Fellowship
and the 2026 Passport AI Leaders Grant. No additional external funding was
received specifically for this work.

\paragraph*{Competing interests.}
The authors declare no relevant financial or non-financial competing interests.

\paragraph*{Code availability.}
Public-data analysis code and documentation will be released upon acceptance.

\paragraph*{AI assistance.}
ChatGPT (OpenAI) and OpenAI Codex assisted with drafting and revising the
abstract and Sections~1--4, preparing Figure~1 and Tables~1--3, and developing
LaTeX, analysis code, and validation scripts. The authors specified the study
design, executed and verified all analyses, reviewed all AI-assisted content,
and take responsibility for the manuscript, code, and results. Neither system
was an evaluated scoring interface, directly accessed study data, or
independently produced clinical labels, model predictions, or reported
numerical values.

%% file: references.bib
@article{men2020oarqa,
  title   = {Automated Quality Assurance of {OAR} Contouring for Lung
             Cancer Based on Segmentation With Deep Active Learning},
  author  = {Men, Kuo and Geng, Huaizhi and Biswas, Tithi and
             Liao, Zhongxing and Xiao, Ying},
  journal = {Frontiers in Oncology},
  volume  = {10},
  year    = {2020},
  doi     = {10.3389/fonc.2020.00986},
  url       = {https://www.frontiersin.org/journals/oncology/articles/10.3389/fonc.2020.00986/full},
  note         = {Art. no. 986}
}

@article{isaksson2022contourqa,
  title   = {Quality Assurance for Automatically Generated Contours
             with Additional Deep Learning},
  author  = {Isaksson, Lars Johannes and Summers, Paul and
             Bhalerao, Abhir and others},
  journal = {Insights into Imaging},
  volume  = {13},
  number  = {1},
  year    = {2022},
  doi     = {10.1186/s13244-022-01276-7},
  url       = {https://link.springer.com/article/10.1186/s13244-022-01276-7},
  note         = {Art. no. 137}
}

@article{luan2024vlmcontourqa,
  title   = {A Multi-Modal Vision-Language Pipeline Strategy for Contour
             Quality Assurance and Adaptive Optimization},
  author  = {Luan, Shunyao and Ou-Yang, Jun and Yang, Xiaofei and
             Wei, Wei and Xue, Xudong and Zhu, Benpeng},
  journal = {Physics in Medicine \& Biology},
  volume  = {69},
  number  = {6},
  year    = {2024},
  doi     = {10.1088/1361-6560/ad2a97},
  url     = {https://doi.org/10.1088/1361-6560/ad2a97},
  note         = {Art. no. 065005}
}

@inproceedings{chen2026segae,
  title     = {Large-Scale Label Quality Assessment for Medical
               Segmentation via a Vision-Language Judge and Synthetic Data},
  author    = {Chen, Yixiong and Zhou, Zongwei and Li, Wenxuan
               and Yuille, Alan L.},
  booktitle = {2026 {IEEE} 23rd International Symposium on Biomedical
               Imaging ({ISBI})},
  pages     = {1--5},
  month     = {April},
  year      = {2026},
  address   = {London, United Kingdom},
  publisher = {IEEE},
  doi       = {10.1109/ISBI61048.2026.11515717},
  url          = {https://doi.org/10.1109/ISBI61048.2026.11515717},
}

@inproceedings{hewitt2019probes,
  title     = {Designing and Interpreting Probes with Control Tasks},
  author    = {Hewitt, John and Liang, Percy},
  booktitle = {Proceedings of the 2019 Conference on Empirical Methods
               in Natural Language Processing and the 9th
               International Joint Conference on Natural Language
               Processing ({EMNLP-IJCNLP})},
  pages     = {2733--2743},
  month     = {November},
  year      = {2019},
  address   = {Hong Kong, China},
  publisher = {Association for Computational Linguistics},
  doi       = {10.18653/v1/D19-1275},
  url       = {https://aclanthology.org/D19-1275/}
}

@article{belinkov2022probes,
  title     = {Probing Classifiers: Promises, Shortcomings, and Advances},
  author    = {Belinkov, Yonatan},
  journal   = {Computational Linguistics},
  volume    = {48},
  number    = {1},
  pages     = {207--219},
  year      = {2022},
  publisher = {MIT Press},
  doi       = {10.1162/coli_a_00422},
  url       = {https://aclanthology.org/2022.cl-1.7/}
}

@inproceedings{nourbakhsh2026medprob,
  title     = {{MedProb}: Probing Internal Representations of
               Vision-Language Models for Medical Question Answering},
  author    = {Nourbakhsh, Erfan and Yang, Ke and Rios, Anthony},
  booktitle = {Findings of the Association for Computational Linguistics:
               {EMNLP} 2026},
  pages     = {27961--27999},
  month     = {October},
  year      = {2026},
  address   = {Budapest, Hungary},
  publisher = {Association for Computational Linguistics},
  doi       = {10.18653/2026.findings-emnlp.1320},
  url       = {https://aclanthology.org/2026.findings-emnlp.1320/}
}

@article{wang2026underutilized,
  title         = {How Medical {VLMs} Underutilize Their Vision Encoders:
                   A Dermatology Perspective},
  author        = {Wang, Janet and Zhang, Yunbei and Wang, Xiao and Hamm, Jihun},
  journal       = {arXiv preprint arXiv:2609.36557},
  year          = {2026},
  eprint        = {2609.36557},
  archivePrefix = {arXiv},
  doi           = {10.48550/arXiv.2609.36557},
  url           = {https://arxiv.org/abs/2609.36557}
}

@article{sellergren2026medgemma15,
  title         = {{MedGemma} 1.5 Technical Report},
  author        = {Sellergren, Andrew and Gao, Chufan and Mahvar, Fereshteh and others},
  journal       = {arXiv preprint arXiv:2604.05081},
  year          = {2026},
  eprint        = {2604.05081},
  archivePrefix = {arXiv},
  url       = {https://arxiv.org/abs/2604.05081v2},
  doi           = {10.48550/arXiv.2604.05081},
  note         = {Version 2, revised May 1, 2026},
}

@inproceedings{loshchilov2019decoupled,
  title     = {Decoupled Weight Decay Regularization},
  author    = {Loshchilov, Ilya and Hutter, Frank},
  booktitle = {International Conference on Learning Representations},
  year      = {2019},
  url       = {https://openreview.net/forum?id=Bkg6RiCqY7}
}

@article{saito2015precisionrecall,
  title   = {The Precision--Recall Plot Is More Informative than the
             {ROC} Plot When Evaluating Binary Classifiers on
             Imbalanced Datasets},
  author  = {Saito, Takaya and Rehmsmeier, Marc},
  journal = {PLOS ONE},
  volume  = {10},
  number  = {3},
  year    = {2015},
  doi     = {10.1371/journal.pone.0118432},
  url       = {https://journals.plos.org/plosone/article?id=10.1371/journal.pone.0118432},
  note         = {Art. no. e0118432}
}

@article{field2007clusterbootstrap,
  title   = {Bootstrapping Clustered Data},
  author  = {Field, C. A. and Welsh, A. H.},
  journal = {Journal of the Royal Statistical Society: Series B
             (Statistical Methodology)},
  volume  = {69},
  number  = {3},
  pages   = {369--390},
  year    = {2007},
  doi     = {10.1111/j.1467-9868.2007.00593.x},
  url     = {https://doi.org/10.1111/j.1467-9868.2007.00593.x}
}

@article{podobnik2023hanseg,
  title   = {{HaN-Seg}: The Head and Neck Organ-at-Risk {CT} and {MR}
             Segmentation Dataset},
  author  = {Podobnik, Ga{\v{s}}per and Strojan, Primo{\v{z}} and
             Peterlin, Primo{\v{z}} and Ibragimov, Bulat and
             Vrtovec, Toma{\v{z}}},
  journal = {Medical Physics},
  volume  = {50},
  number  = {3},
  pages   = {1917--1927},
  year    = {2023},
  doi     = {10.1002/mp.16197},
  url     = {https://doi.org/10.1002/mp.16197}
}

@article{luo2023segrap2023,
  title   = {{SegRap2023}: A Benchmark of Organs-at-Risk and Gross
             Tumor Volume Segmentation for Radiotherapy Planning of
             Nasopharyngeal Carcinoma},
  author  = {Luo, Xiangde and Fu, Jia and Zhong, Yunxin and others},
  journal = {Medical Image Analysis},
  volume  = {101},
  year    = {2025},
  doi     = {10.1016/j.media.2024.103447},
  url     = {https://doi.org/10.1016/j.media.2024.103447},
  note    = {Art. no. 103447}
}
